\documentclass[usenatbib,usegraphicx]{mn2e}
\def\msun{{\rm M}_{\odot}}
\def\rsun{{\rm R}_{\odot}}

\def\spitzer{{\it Spitzer\ }}

\begin{document}

\title[Rotation of low-mass stars in NGC 2362]{The Monitor project:
    Rotation of low-mass stars in NGC 2362 -- testing the disc regulation paradigm at 5 Myr}
\author[J.~M.~Irwin et al.]{Jonathan~Irwin$^{1,2}$\thanks{E-mail: jmi at
    ast.cam.ac.uk}, Simon~Hodgkin$^{1}$,
    Suzanne~Aigrain$^{3}$, Jerome~Bouvier$^{4}$,
\newauthor
Leslie~Hebb$^{5}$, Mike~Irwin$^{1}$, Estelle~Moraux$^{4}$ \\
$^{1}$Institute of Astronomy, University of Cambridge, Madingley Road,
  Cambridge, CB3 0HA, United Kingdom \\
$^{2}$Harvard-Smithsonian Center for Astrophysics, 60 Garden Street,
    Cambridge, MA 02138, USA \\
$^{3}$Astrophysics Group, School of Physics, University of Exeter, Stocker Road,
  Exeter, EX4 4QL, United Kingdom \\
$^{4}$Laboratoire d'Astrophysique, Observatoire de Grenoble, BP 53,
  F-38041 Grenoble C\'{e}dex 9, France \\
$^{5}$School of Physics and Astronomy, University of St Andrews,
  North Haugh, St Andrews, KY16 9SS, Scotland}
\date{}

\maketitle

\begin{abstract}
We report on the results of a time-series photometric survey of NGC
2362, carried out using the CTIO 4m Blanco telescope and Mosaic-II
detector as part of the Monitor project.  Rotation periods were
derived for $271$ candidate cluster members over the mass range $0.1
\la M/\msun \la 1.2$.  The rotation period distributions show a clear
mass-dependent morphology, qualitatively similar to that in NGC 2264,
as would be expected from the age of this cluster.  Using models of
angular momentum evolution, we show that angular momentum losses over
the $\sim 1-5\ {\rm Myr}$ age range appear to be needed in order to
reproduce the evolution of the slowest rotators in the sample from the
ONC to NGC 2362, as found by many previous studies.  By incorporating
\spitzer IRAC mid-IR measurements, we found that $3-4$ objects showing
mid-IR excesses indicative of the presence of circumstellar discs were
all slow rotators, as would be expected in the disc regulation
paradigm for early pre-main sequence angular momentum evolution, but
this result is not statistically significant at present, given the
extremely limited sample size.
\end{abstract}

\section{Introduction}
\label{intro_section}

NGC 2362 is a very young open cluster, at a moderate distance.  We
adopt the parameters of \citet{moi2001} for the remainder of this
work: an age of $5^{+1}_{-2}\ {\rm Myr}$, distance $1480\ {\rm pc}$
and reddening $E(B-V) = 0.10\ {\rm mag}$.  Note however that there is
some controversy in the literature regarding the age of this cluster,
with estimates ranging from $3 - 9\ {\rm Myr}$.  \citet{may07} have
examined the relative ages of young open clusters, finding that NGC
2362 does indeed seem to be older than the ONC, but that it may be
younger than IC 348, and they suggest an age of $\sim 3\ {\rm
  Myr}$.

Nevertheless, the extremely young age, and lack of substantial
nebulosity or differential reddening in this cluster make it an ideal
testing ground for theories of pre-main sequence stellar evolution.
\citet{dahm05} carried out an extensive spectroscopic survey,
concentrating on the H$\alpha$ $6563\ {\rm \AA}$ and Li\ {\sc i}
$6707.8\ {\rm \AA}$ features, finding that a large fraction ($\sim 91$
per cent) of the T Tauri star (TTS) population in NGC 2362 is composed
of weak-line emitters (WTTS), as expected given the canonical $5\ {\rm
Myr}$ age for the cluster.  They found that $\sim 5-9$ per cent of the
TTS population are likely to still be undergoing accretion on the
basis of the measured H$\alpha$ equivalent widths, which is comparable
to the fraction of objects found to exhibit L-band excess emission
indicative of an inner disc by \citet{hll2001}.

NGC 2362 has also been observed more recently with \spitzer using
the IRAC (Infra-Red Array Camera) instrument in the $3.6$, $4.5$,
$5.8$ and $8.0\ {\rm \mu m}$ bands.  \citet{dh07} used these
observations to derive spectral energy distributions for candidate NGC
2362 members, and to characterise disc emission in objects exhibiting
infra-red excesses.  They found an upper limit to the fraction of
objects exhibiting optically thick discs of $\sim 7 \pm 2$ per cent,
with an additional $\sim 12 \pm 3$ per cent of objects exhibiting weak
or optically thin discs.

\subsection{Evolution of stellar angular momentum}
\label{amevol_section}

A long-standing issue in our understanding of the evolution of stellar
rotation rates on the pre-main sequence is that significant loss of
angular momentum appears to be required in order to reproduce the
observed rotation rates of low-mass stars.

During the first few Myr of their evolution, low-mass pre-main
sequence (PMS) stars contract by factors of $\sim 2-3$.  Hence, if
angular momentum is conserved, all sub-solar mass stars, which are
thought to have initial rotation periods of $\sim 10\ {\rm days}$,
should rotate at periods shorter than $2\ {\rm days}$ after only $2-3\
{\rm Myr}$.  Observations at these ages show that some stars rotate
much slower than this rate, and indeed the slowest rotators do not
appear to spin up at all over the first few Myr of their evolution.

The most popular scenarios for solving this problem invoke additional
angular momentum losses resulting from the interaction of the star
with a circumstellar disc (e.g. \citealt{k91}).  In the ``disc
locking'' scenario, this interaction results in transfer of angular
momentum from the star to the disc, which maintains a constant stellar 
rotation rate until the disc dissipates sufficiently to break the
coupling.  Alternatively, the angular momentum may be lost through an
enhanced stellar wind, powered by the accretion of material from the
disc \citep{mp05}.

In both cases, the most basic prediction of the models is that there
should be a correlation between rotation rate and the presence of
discs, in the sense that the slow rotators should still be surrounded
by discs, whereas for the rapid rotators, the discs should have
dispersed.

Much of the recent literature on rotation has been devoted to the
search for a correlation between rotation rate and disc indicators,
including diagnostics of accretion (UV excess, spectral line
diagnostics such as H$\alpha$, etc.) and of the thermal emission from
the disc itself (infra-red excess).  Many of the early studies were
inconclusive, due to a combination of effects including small sample
sizes, biases, dependence of the adopted disc indicators on other
stellar parameters (predominantly mass), and the use of ambiguous disc
indicators.

Two recent studies (\citealt{rebull06}; \citealt{cb07}) have claimed
statistically significant detections of the expected correlation
between slow rotators and circumstellar discs, in the ONC ($1 \pm 1\
{\rm Myr}$; \citealt{h97}) and NGC 2264 ($2-4\ {\rm Myr}$;
\citealt{park2000}), respectively.  \citet{cb06} have also examined
IC 348 ($2-3\ {\rm Myr}$; \citealt{luh2003}), but the results were
ambiguous due to small number statistics.  These studies have all used
the IRAC instrument on the \spitzer space telescope, which provides
mid-IR photometry in four broad bands: $3.6$, $4.5$, $5.8$ and $8.0\
\micron$.  In particular, $8.0\ \micron$ infra-red flux excesses
provide a much more reliable indicator of the presence of
circumstellar discs than shorter-wavelength IR excesses (e.g. in the
$K$-band).  There is thus mounting evidence in favour of the disc
regulation paradigm, but nevertheless, it is important to verify that
this result extends to older open clusters.

\subsection{The survey}

We have undertaken a photometric survey in NGC 2362 using the CTIO 4m
Blanco telescope and Mosaic-II imager.  Our goals are two-fold: first,
to study rotation periods for a sample of low-mass members, covering
K and M spectral types, down to $\sim 0.1\ \msun$, and second, to
look for eclipsing binary systems containing low-mass stars, to obtain
dynamical mass measurements in conjunction with radial velocities from
follow-up spectroscopy.  Such systems provide the most accurate
determinations of fundamental stellar parameters (in particular,
masses) for input to models of stellar evolution, which are poorly
constrained in this age range.  We defer discussion of our search for
occultations to another publication (Miller et al., in preparation).

These observations are part of a larger photometric monitoring
survey of young open clusters over a range of ages and metalicities
(the Monitor project; \citealt{hodg06} and \citealt{a2007}).

The remainder of the paper is structured as follows: the observations
and data reduction are described in \S \ref{odr_section}, and the
colour magnitude diagram (CMD) of the cluster and candidate membership
selection are presented in \S \ref{memb_section}.  The method we use
for obtaining photometric periods is summarised in \S
\ref{period_section} (see \citealt{i2006} for a more detailed
discussion).  Our results are given in \S \ref{results_section}, and
\S \ref{conclusions_section} summarises our conclusions.

In several of the following sections, we use mass and radius estimates
for the cluster members.  These were derived from the $5\ {\rm Myr}$
NextGen models, using the $I$-band magnitudes, rather than $V - I$
colour or $V$ magnitude, for reasons discussed in \S
\ref{cmd_section}.

\section{Observations and data reduction}
\label{odr_section}

Photometric monitoring observations were obtained using the 4m CTIO
Blanco telescope, with the Mosaic-II imager, on $8 \times 1/2$--nights
between 2005 Dec 24 and 2006 Jan 06.  This instrument provides a field
of view of $\sim 36' \times 36'$ ($0.37\ {\rm sq. deg}$), using a
mosaic of eight ${\rm 2k} \times {\rm 4k}$ pixel CCDs, at a scale of
$\sim 0.27'' / {\rm pix}$.

\begin{figure}
\centering
\includegraphics[angle=0,width=3.2in]{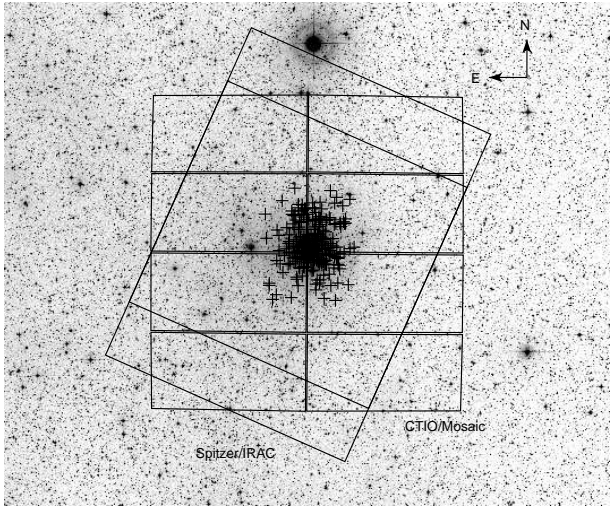}

\caption{Digitised sky survey (DSS) image of NGC 2362 covering $\sim
  1.2^\circ \times 1^\circ$, showing the coverage of the present
  survey (8-chip mosaic tile), and the \spitzer IRAC field of
  \citet{dh07}, showing the total coverage (larger box), and the
  smaller central region covered by all four bands.  Crosses show the
  positions of the candidate cluster members from \citet{dh07}.}

\label{coverage}
\end{figure}

A single field in NGC 2362, centred on the star $\tau$ CMa (see Figure
\ref{coverage}), was observed for $\sim 4\ {\rm hours}$ per night, in
parallel with another field in the cluster M50, the results from which
will be published separately.  Exposure times were $75\ {\rm s}$ in
$i$-band, giving a cadence of $\sim 6\ {\rm minutes}$ (composed of $2
\times$ $75\ {\rm s}$ exposures plus $2 \times$ $100\ {\rm s}$ readout
time, slewing between M50 and NGC 2362 during readout).  We also
obtained deep $V$-band exposures ($600\ {\rm s}$, $450\ {\rm s}$, $2
\times 300\ {\rm s}$ and $150\ {\rm s}$) which were stacked and used
to produce a colour-magnitude diagram of the cluster.

\begin{figure}
\centering
\includegraphics[angle=270,width=3.2in]{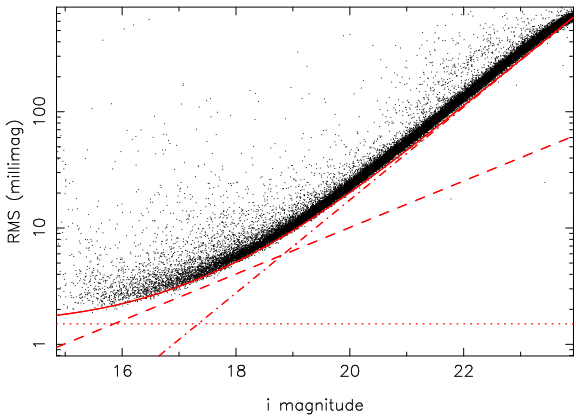}

\caption{Plot of RMS scatter per data point (measured over the entire
  data-set of $8 \times$ $1/2$-nights) as a function of magnitude
  for the $i$-band observations of a single field in NGC 2362, for all
  unblended objects with stellar morphological classifications.  The
  diagonal dashed line shows the expected RMS from Poisson noise in
  the object, the diagonal dot-dashed line shows the RMS from sky
  noise in the photometric aperture, and the dotted line shows an
  additional $1.5\ {\rm mmag}$ contribution added in quadrature to
  account for systematic effects.  The solid line shows the overall
  predicted RMS, combining these contributions.}

\label{rmsplot}
\end{figure}

For a full description of our data reduction steps, the reader is
referred to \citet{i2007a}.  Briefly, we used the pipeline for
the INT wide-field survey \citep{il2001} for 2-D 
instrumental signature removal (bias correction, flatfielding,
defringing) and astrometric and photometric calibration.  We then 
generated a master catalogue for each filter by stacking $20$
of the frames taken in the best conditions (seeing, sky brightness and
transparency) and running the source detection software on the stacked
image.  The resulting source positions were used to perform aperture
photometry on all of the time-series images.  We achieved a per data
point photometric precision of $\sim 2-4\ {\rm mmag}$ for the
brightest objects, with RMS scatter $< 1$ per cent for $i \la 19$ (see
Figure \ref{rmsplot}).  A signal-to-noise ratio of $5$ (corresponding
approximately to the detection limit for point sources on a single
frame of the differential photometry) is reached at $i \sim 22.7$.

Our source detection software flags any objects detected as having
overlapping isophotes.  This information is used, in conjunction with
a morphological image classification flag also generated by the
pipeline software \citep{il2001} to allow us to identify non-stellar
or blended objects in the time-series photometry.

Photometric calibration of our data was carried out using regular
observations of \citet{l92} equatorial standard star fields in the
usual way.

Light curves were extracted from the data for $\sim 85\,000$ objects,
$56\,000$ of which had stellar morphological classifications ($\sim
40$ per cent of these are flagged as blended), using our standard
aperture photometry techniques, described in \citet{i2007a}. We fit a
2-D quadratic polynomial to the residuals in each frame  (measured for
each object as the difference between its magnitude on the frame in
question and the median calculated across all frames) as a function of
position, for each of the $8$ CCDs separately. Subsequent removal of
this function accounts for effects such as varying differential
atmospheric extinction across each frame.  Over a single CCD, the
spatially-varying part of the correction remains small, typically
$\sim 0.02\ {\rm mag}$ peak-to-peak.  The reasons for using this
technique are discussed in more detail in \citet{i2007a}.

For the production of deep CMDs, we stacked 20 $i$-band observations,
taken in good seeing and photometric conditions, and all of the
$V$-band observations.  The limiting magnitudes on these stacked
images, measured as the approximate magnitude at which our catalogues
are $50$ per cent complete, were $V \simeq 24.4$ and $i \simeq 23.6$.

\section{Selection of candidate low-mass members}
\label{memb_section}

Catalogues of candidate NGC 2362 members are available in the
literature (\citealt{moi2001}; \citealt{dahm05}), but we elected to
perform a new photometric selection using $V$ versus $V-I$ CMDs from
our data in order to properly match the relatively large field of view
of our time-series observations compared to these previous studies.

\subsection{The $V$ versus $V - I$ CMD}
\label{cmd_section}

\begin{figure}
\centering
\includegraphics[angle=270,width=3.5in]{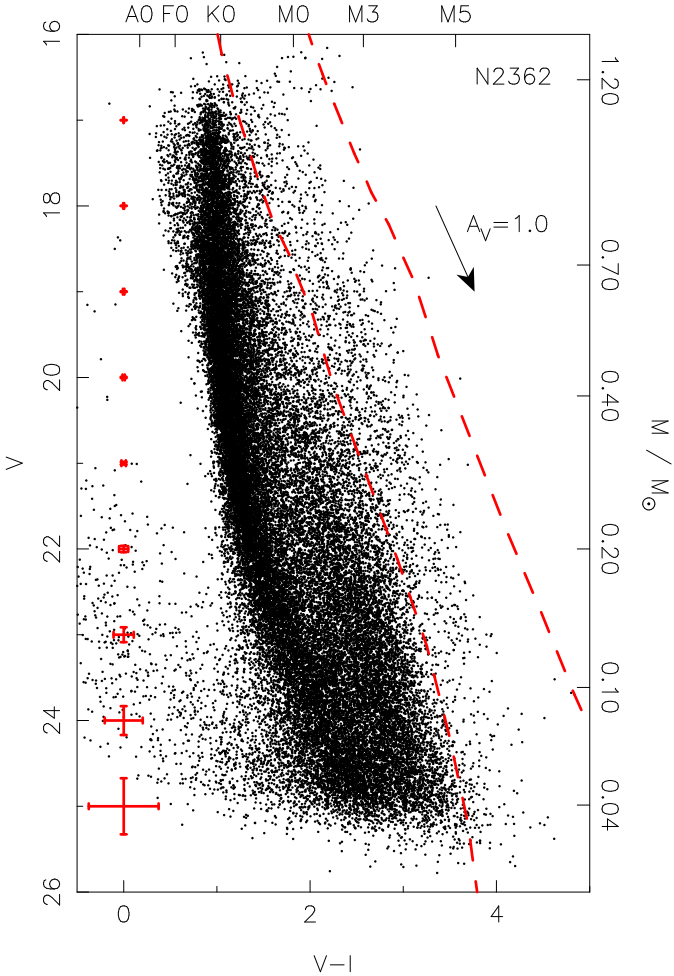}

\caption{$V$ versus $V - I$ CMD of NGC 2362 from stacked images, for
all objects with stellar morphological classification.  The cluster
sequence is clearly visible on the right-hand side of the diagram.
The boundaries of the region used to select photometric candidate
members are shown by the dashed lines (all objects between the dashed
lines were selected).  The reddening vector for $A_V = 1.0$ is shown
at the right-hand side of the diagram.  The mass scale is from the $5\
{\rm Myr}$ NextGen models \citep{bcah98} for $M > 0.1\ \msun$, and the
$5\ {\rm Myr}$ DUSTY models \citep{cbah2000} for $M < 0.1\ \msun$,
using our empirical isochrone to convert the $V$ magnitudes to $I$
magnitudes, and subsequently obtaining the masses from these, due to
known problems with the $V$ magnitudes from the models (see \S
\ref{cmd_section}).  The error bars at the left-hand side of the plot
indicate the typical photometric error for an object on the cluster
sequence.}

\label{cmd}
\end{figure}

Our CMD of NGC 2362 is shown in Figure \ref{cmd}.  The $V$ and $i$
measurements were converted to the standard Johnson-Cousins
photometric system using colour equations derived from our standard
star observations:
\begin{eqnarray}
(V - I)& = &(V_{ccd} - i_{ccd})\ /\ 0.899 \\
V& = &V_{ccd} + 0.005\ (V - I) \\
I& = &i_{ccd} - 0.096\ (V - I)
\end{eqnarray}

Candidate cluster members were selected by defining an empirical cluster
sequence `by eye' to follow the clearly-visible cluster single-star
sequence.  The cuts were defined by moving this line along a vector
perpendicular to the cluster sequence, by amounts $k - \sigma(V -
I)$ and $k + \sigma(V - I)$ as measured along this vector, where
$\sigma(V - I)$ is the photometric error in the $V - I$ colour.  The
values of $k$ used were $-0.25\ {\rm mag}$ for the lower line and
$0.7\ {\rm mag}$ for the upper line on the diagram, making the
brighter region wider to avoid rejecting binary and multiple systems,
which are overluminous for their colour compared to single stars.
$1800$ candidate photometric members were selected, over the full $V$
magnitude range from $V = 16$ to $26$, but the well-defined cluster
sequence appears to terminate at $M \sim 0.1\ \msun$, or $V \sim
23.5$, with a few candidate brown dwarfs found below this limit, but
with high field contamination.

We also considered using the model isochrones of \citet{bcah98} and
\citet{cbah2000} for selecting candidate members.  The NextGen model
isochrones were found to be unsuitable due to the known discrepancy
between these models and observations in the $V - I$ colour for
$T_{\rm eff} \la 3700\ {\rm K}$ (corresponding here to $V - I \ga 2$).
This was examined in more detail by \citet{bcah98}, and is due to a
missing source of opacity at these temperatures, leading to
overestimation of the $V$-band flux.  Consequently, when we have used
the NextGen isochrones to determine model masses and radii for our
objects, the $I$-band absolute magnitudes were used to perform the
relevant look-up, since these are less susceptible to the missing
source of opacity, and hence give more robust estimates.

\subsection{Contamination}
\label{contam_section}

Since the cluster sequence is not well-separated from the field in the
CMD, it is important to estimate the level of field star contamination
in the sample of candidate cluster members.  This has been done using
two independent methods.

\subsubsection{Galactic models}
\label{galmod_sect}

\begin{figure}
\centering
\includegraphics[angle=270,width=3in]{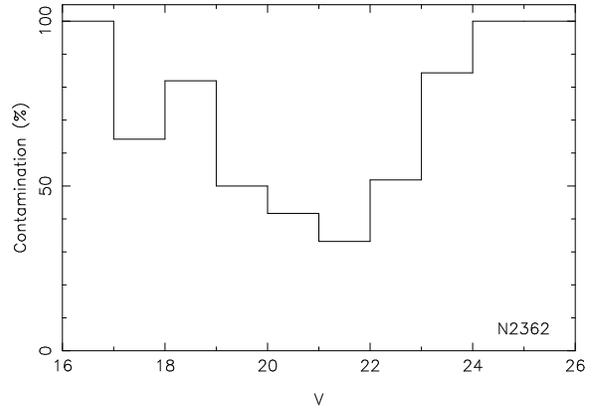}

\caption{Contamination estimated from Galactic models, measured as the
  ratio of the calculated number of objects in each magnitude bin from
  the models, to the number of objects detected and classified as
  candidate cluster members in that magnitude bin.  Note that bins
  with contamination estimates $> 100$ per cent (where there were more
  objects in that bin from the Galactic model than were actually
  observed) have been truncated to $100$ per cent.}

\label{contam}
\end{figure}

The Besan\c{c}on Galactic models \citep{r2003} were used to to
generate a simulated catalogue of objects passing our selection
criteria at the Galactic coordinates of NGC 2362 ($l = 238.2^\circ$,
$b = -5.5^\circ$), covering the total FoV of $\sim 0.35\ {\rm sq.deg}$
(including gaps between detectors).  The same selection process as
above for the cluster members was applied to this catalogue to find
the contaminant objects.  A total of $1500$ simulated objects passed
these membership selection criteria, giving an overall contamination
level of $\sim 65$ per cent after correcting for bins where the number
of objects predicted by the models exceeded the number actually
observed by $\sim 20$ per cent (we simply assumed $100$ per cent field
contamination in these bins). Figure \ref{contam} shows the
contamination as a function of $V$ magnitude.  Note that this figure
is somewhat uncertain due to the need to use Galactic models, and
especially given the overestimation of the numbers of observed objects
by the models, which lends low confidence to the results derived
in this section.  We therefore pursue an alternative method for
obtaining a limit to the contamination level below.

\subsubsection{Radial distribution}

An alternative method for generating a simple upper limit to the
contamination level has also been applied to verify the results
obtained from the Galactic models.  Since the cluster is relatively
centrally concentrated, we divided the population of candidate cluster
members into two bins in radius, using as the centre the position of
the star $\tau$ CMa.  Number counts of objects selected using the CMD
as candidate members were then produced inside and outside of a $15'$
radius, chosen since it roughly divides the area covered into two,
providing good statistics in both bins.  By assuming the stars inside
$15'$ are cluster members, and those outside are not, we can obtain a
simple upper limit to the contamination level by dividing these
counts, and correcting for the relative area covered by each bin.

\begin{figure}
\centering
\includegraphics[angle=270,width=3in]{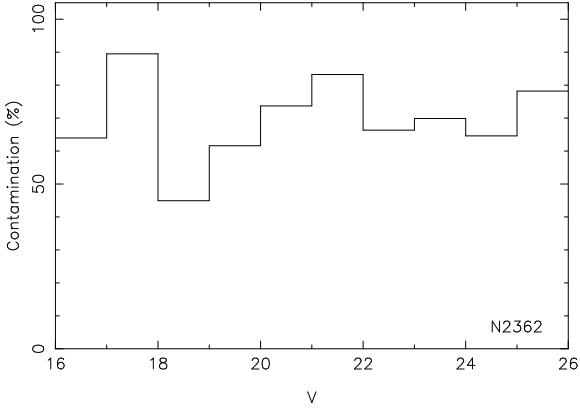}

\caption{Upper limit to the field contamination level estimated from
  the radial distribution method (see text), plotted as a function of
  $V$-band magnitude.}
\label{contam_radial}
\end{figure}

Figure \ref{contam_radial} shows the resulting distribution of
contamination as a function of magnitude, for comparison with the
Galactic model results in Figure \ref{contam}.  The overall
contamination level integrated over all magnitude bins from this
method is $68$ per cent, comparable to the estimate using the Galactic
models from Section \ref{galmod_sect}.

\section{Period detection}
\label{period_section}

\subsection{Method}
\label{method_section}

The method we use for detection of periodic variables is described in
detail in \citet{i2006}, and we provide only a brief
summary here.  The method uses least-squares fitting of sine curves to
the time series $m(t)$ (in magnitudes) for {\em all} candidate cluster
members, using the form:
\begin{equation}
m(t) = m_{dc} + \alpha \sin(\omega t + \phi)
\label{sine_eqn}
\end{equation}
where $m_{dc}$ (the DC light curve level), $\alpha$ (the amplitude) and
$\phi$ (the phase) are free parameters at each value of $\omega$ over
an equally-spaced grid of frequencies, corresponding to periods from
$0.005 - 50\ {\rm days}$ for the present data-set.

Periodic variable light curves were selected by evaluating the change
in reduced $\chi^2$:
\begin{equation}
\Delta \chi^2_\nu = \chi^2_\nu - \chi^2_{\nu,{\rm smooth}} > 0.4
\end{equation}
where $\chi^2_\nu$ is the reduced $\chi^2$ of the original light curve
with respect to a constant model, and $\chi^2_{\nu,{\rm smooth}}$ is
the reduced $\chi^2$ of the light curve with the smoothed, phase-folded
version subtracted.  This threshold was used for the M34 data and
appears to work well here too, carefully checked by examining all the
light curves for two of the detectors, chosen randomly.  A total of
$1008$ objects were selected by this automated part of the procedure.

The selected light curves were examined by eye, to define the final
sample of periodic variables.  A total of $271$ light curves were
selected, with the remainder appearing non-variable or too ambiguous
to be included.

\subsection{Simulations}
\label{sim_section}

Monte Carlo simulations were performed following the method detailed
in \citet{i2006}, injecting simulated signals of $2$ per cent
amplitude and periods chosen following a uniform distribution on
$\log_{10}$ period from $0.1$ to $20\ {\rm days}$, into light curves
covering a uniform distribution in mass, from $1.2$ to $0.1\ \msun$.
A total of $1015$ objects were simulated.

\begin{figure*}
\centering
\includegraphics[angle=270,width=6in]{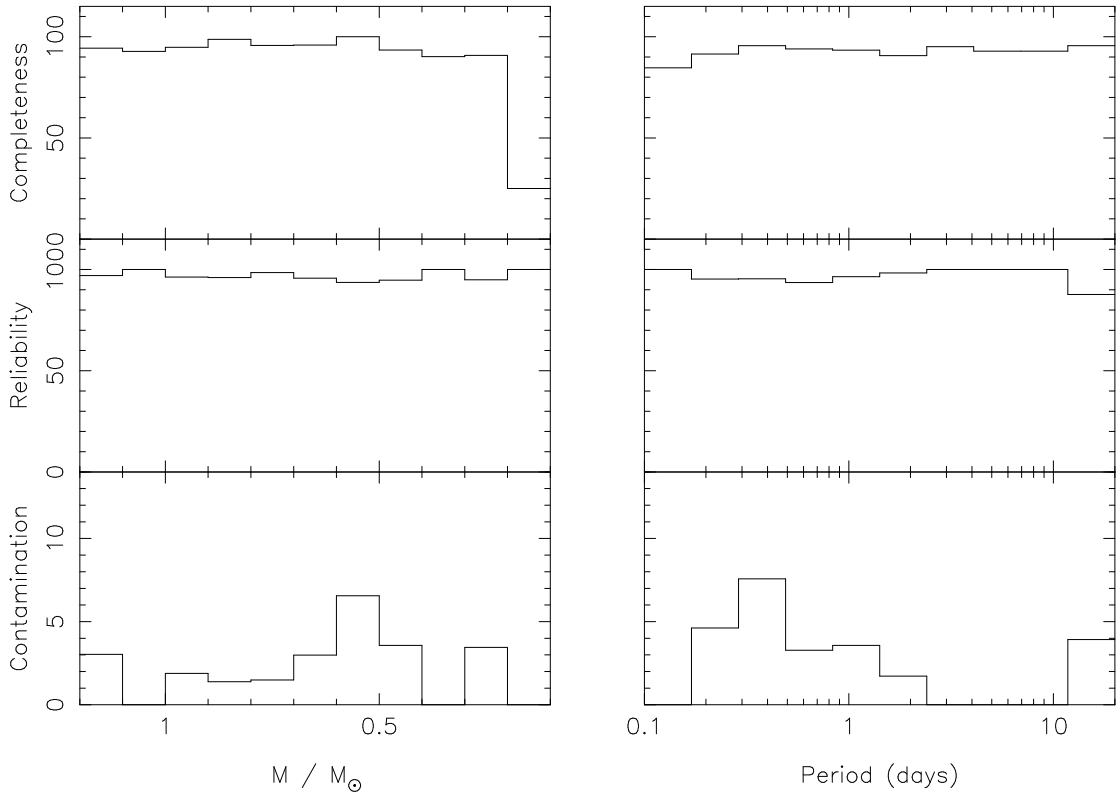}

\caption{Results of the simulations for $0.02\ {\rm mag}$ amplitude
  expressed as percentages, plotted as a function of mass (left) and
  period (right).  The simulated region covered $0.1 < {\rm M}/\msun <
  1.2$ in order to be consistent with the NGC 2362 sample.  {\bf Top
  panels}: completeness as a function of real (input) period.  {\bf
  Centre panels}: Reliability of period determination, plotted as the
  fraction of objects with a given true period, detected with the
  correct period (defined as differing by $< 20$ per cent from the
  true period).  {\bf Bottom panels}: Contamination, plotted as the
  fraction of objects with a given detected period, having a true
  period differing by $> 20$ per cent from the detected value.}

\label{sim_results}
\end{figure*}

\begin{figure}
\centering
\includegraphics[angle=270,width=3in,bb=59 107 581 630,clip]{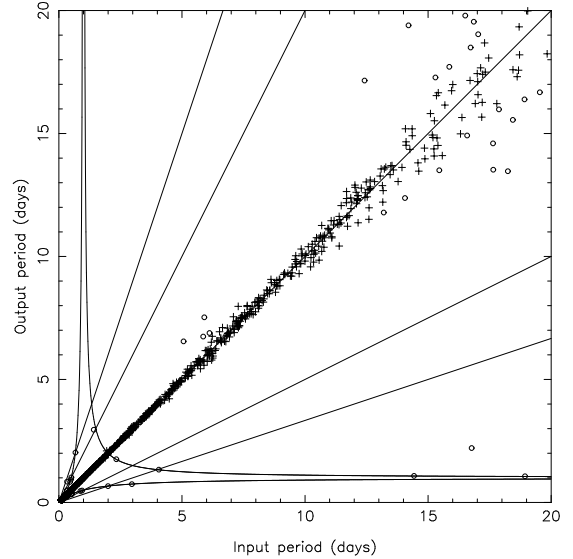}

\caption{Detected period as a function of actual (input) period for our
  simulations.  Objects plotted with crosses had fractional period
  error $< 10$ per cent, open circles $> 10$ per cent.  The straight
  lines represent equal input and output periods, and factors of $2$,
  $3$, $1/2$ and $1/3$.  The curved lines are the loci of the $\pm 1\
  {\rm day^{-1}}$ aliases resulting from gaps during the day.  The
  majority of the points fall on (or close to) the line of equal
  periods.}

\label{periodcomp}
\end{figure}

The results of the simulations are shown in Figure \ref{sim_results}
as diagrams of completeness, reliability and contamination
as a function of period and stellar mass.  Broadly, our period
detections are close to $100$ per cent complete from $1.2\ \msun$ down
to $0.2\ \msun$, with remarkably little period dependence.
Figure \ref{periodcomp} shows a comparison of the detected periods
with real periods for our simulated objects, indicating good
reliability.

\subsection{Detection rate and reliability}

\begin{figure}
\centering
\includegraphics[angle=270,width=3.5in]{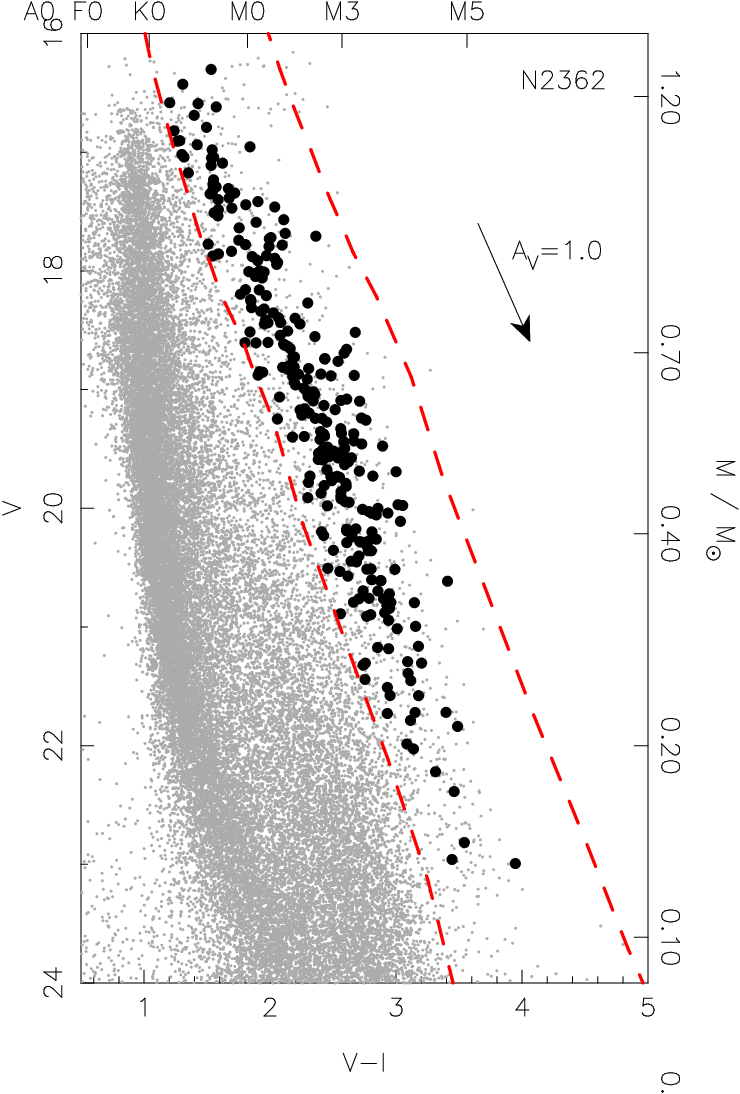}

\caption{Magnified $V$ versus $V - I$ CMD of NGC 2362, for objects
with stellar morphological classification, as Figure \ref{cmd},
showing all $271$ candidate cluster members with detected periods
(black points).  The dashed lines show the cuts used to select
candidate cluster members (see \S \ref{cmd_section}).}

\label{cands_on_cmd}
\end{figure}

The locations of our detected periodic variable candidate cluster
members on a $V$ versus $V-I$ CMD of NGC 2362 are shown in Figure
\ref{cands_on_cmd}.  The diagram indicates that the majority of
the detections lie on the single-star cluster sequence, as would
be expected for rotation in cluster stars as opposed to, say,
eclipsing binaries.

The properties of all our rotation candidates are listed in Table
\ref{cand_table}.

\begin{table*}
\centering
\begin{tabular}{lllrrrrrrrrrrrr}
\hline
Identifier    &RA    &Dec   &$V$   &$I$   &$[3.6]$ &$[4.5]$ &$[5.8]$ &$[8.0]$ &$P$    &$\alpha_i$ &$M$ &$R$ \\
              &J2000 &J2000 &(mag) &(mag) &(mag) &(mag) &(mag) &(mag) &(days) &(mag)      &($\msun$) &($\rsun$) \\
\hline
N2362-1-1357  &07 17 34.82 &-25 10 07.0 &16.90 &15.62 &      &      &      &      & 7.344 &0.005 &1.07 &1.70 \\
N2362-1-2076  &07 17 39.10 &-25 11 32.6 &16.91 &15.64 &      &      &      &      & 7.768 &0.003 &1.06 &1.70 \\
N2362-1-3162  &07 17 48.43 &-25 13 55.3 &21.73 &18.80 &      &      &      &      & 2.463 &0.019 &0.19 &0.73 \\
N2362-1-4646  &07 18 00.08 &-25 08 16.3 &20.52 &17.52 &      &      &      &      & 1.141 &0.015 &0.40 &1.08 \\
N2362-1-4685  &07 18 00.92 &-25 14 24.5 &21.31 &18.10 &      &      &      &      &10.245 &0.014 &0.29 &0.91 \\
\hline
\end{tabular}

\caption{Properties of our $271$ rotation candidates, including $V$
  and $I$-band magnitudes from our CCD photometry, \spitzer IRAC
  magnitudes in the $3.6$, $4.5$, $5.8$ and $8.0\ \micron$ bands,
  where available, the period $P$, $i$-band amplitude $\alpha_i$ (in
  units of magnitudes, in the instrumental bandpass), interpolated
  mass and radius (from the models of \citealt{bcah98}, derived using
  the $I$ magnitudes).  Our identifiers are formed
  using a simple scheme of the cluster name, CCD number
  and a running count of stars in each CCD, concatenated with dashes.
  The full table is available in the electronic edition.  Machine
  readable copies of the data tables from all the Monitor rotation
  period publications are also available at {\tt
  http://www.ast.cam.ac.uk/research/monitor/rotation/}.}
\label{cand_table}
\end{table*}

\section{Results}
\label{results_section}

\subsection{NGC 2362 rotation periods}
\label{prv_section}

\begin{figure}
\centering
\includegraphics[angle=270,width=3in]{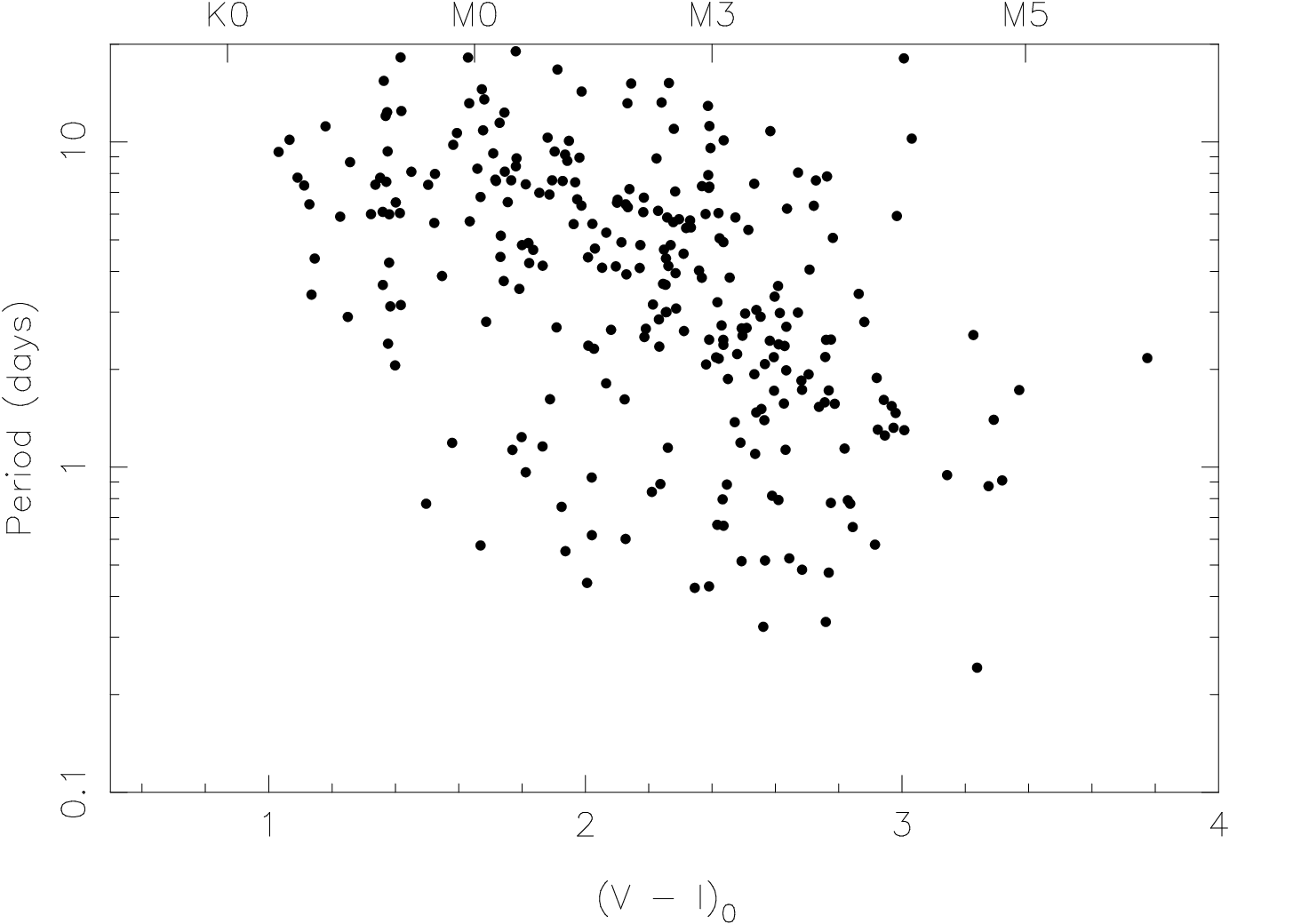}
\includegraphics[angle=270,width=3in]{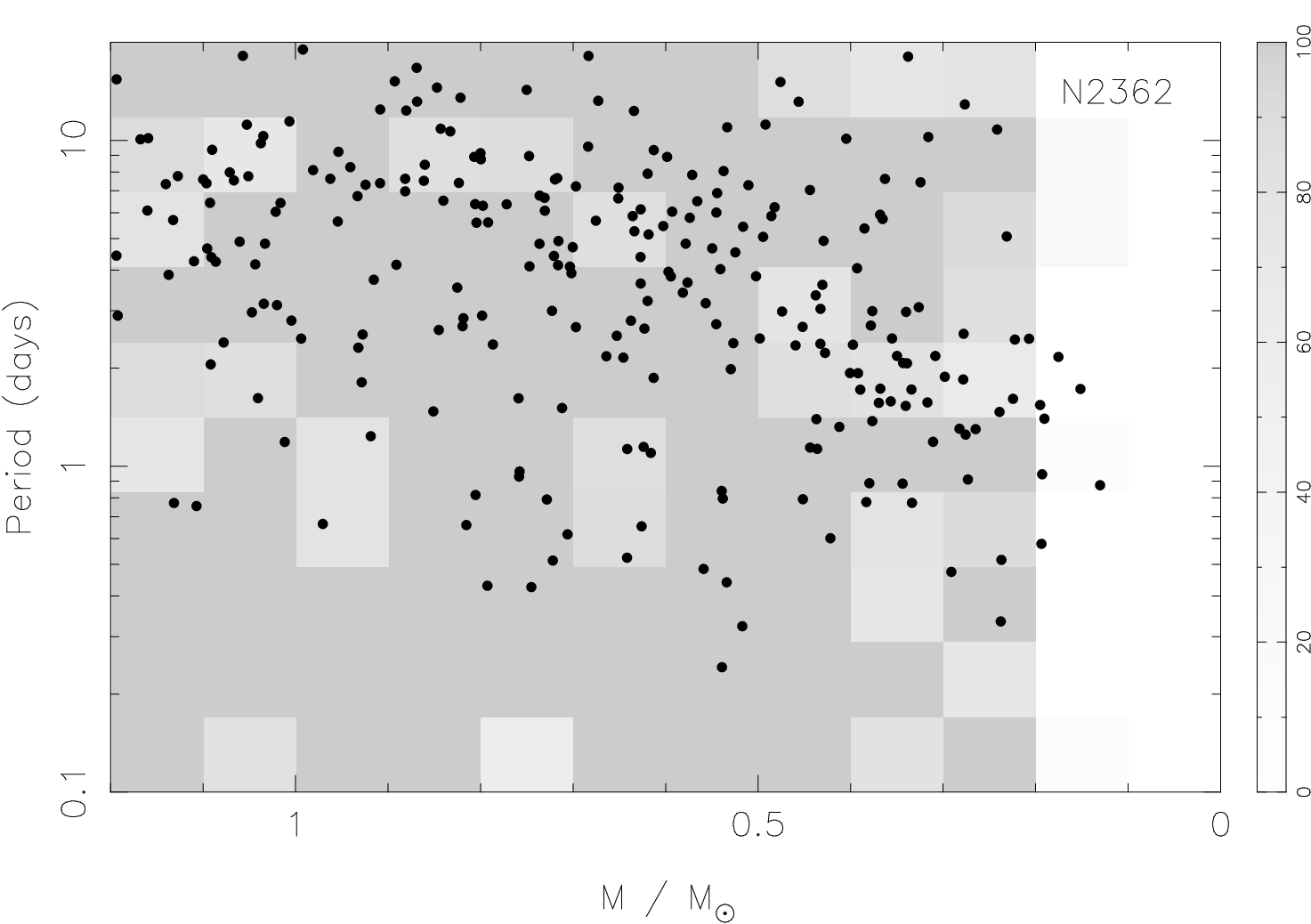}

\caption{Plots of rotation period as a function of dereddened $V-I$
  colour (top), and mass (bottom) for NGC 2362, deriving the masses
  using the $5\ {\rm Myr}$ NextGen mass-magnitude relations of
  \citet{bcah98} and our measured $I$-band magnitudes.  In the lower
  diagram, the greyscales show the completeness for $0.02\ {\rm mag}$
  periodic variations from the simulations.}

\label{pcd}
\end{figure}

Plots of period as a function of $V-I$ colour and mass for the objects
photometrically selected as possible cluster members are shown in
Figure \ref{pcd}.  Below $\sim 0.7\ \msun$ (or M0), these diagrams
reveal a correlation between stellar mass (or spectral type) and the
longest rotation period seen at that mass, with a clear lack of slow
rotators at very low masses.  This trend is also followed by the
majority of the rotators in this mass range, with only a tail of
faster rotators to $\sim 0.5\ {\rm day}$ periods, and very few objects
rotating faster than this.  This is very similar to what we found in
the earlier NGC 2516 and NGC 2547 studies (\citealt{i2007b};
\citealt{i2007d}).

Above $\sim 0.7\ \msun$, the distribution appears to show little mass
dependence, with the slowest rotators all exhibiting $\sim 10\ {\rm
  day}$ periods.

These morphological features do not appear to be a result of sample
biases.  In particular, the simulations show that the survey is
sensitive to much shorter periods than $0.5\ {\rm day}$, and the upper
limit in detectable periods is not mass-dependent, so this cannot
explain the sloping morphology of the upper envelope of rotation
periods in Figure \ref{pcd} at low masses.

\begin{figure}
\centering
\includegraphics[angle=270,width=3in]{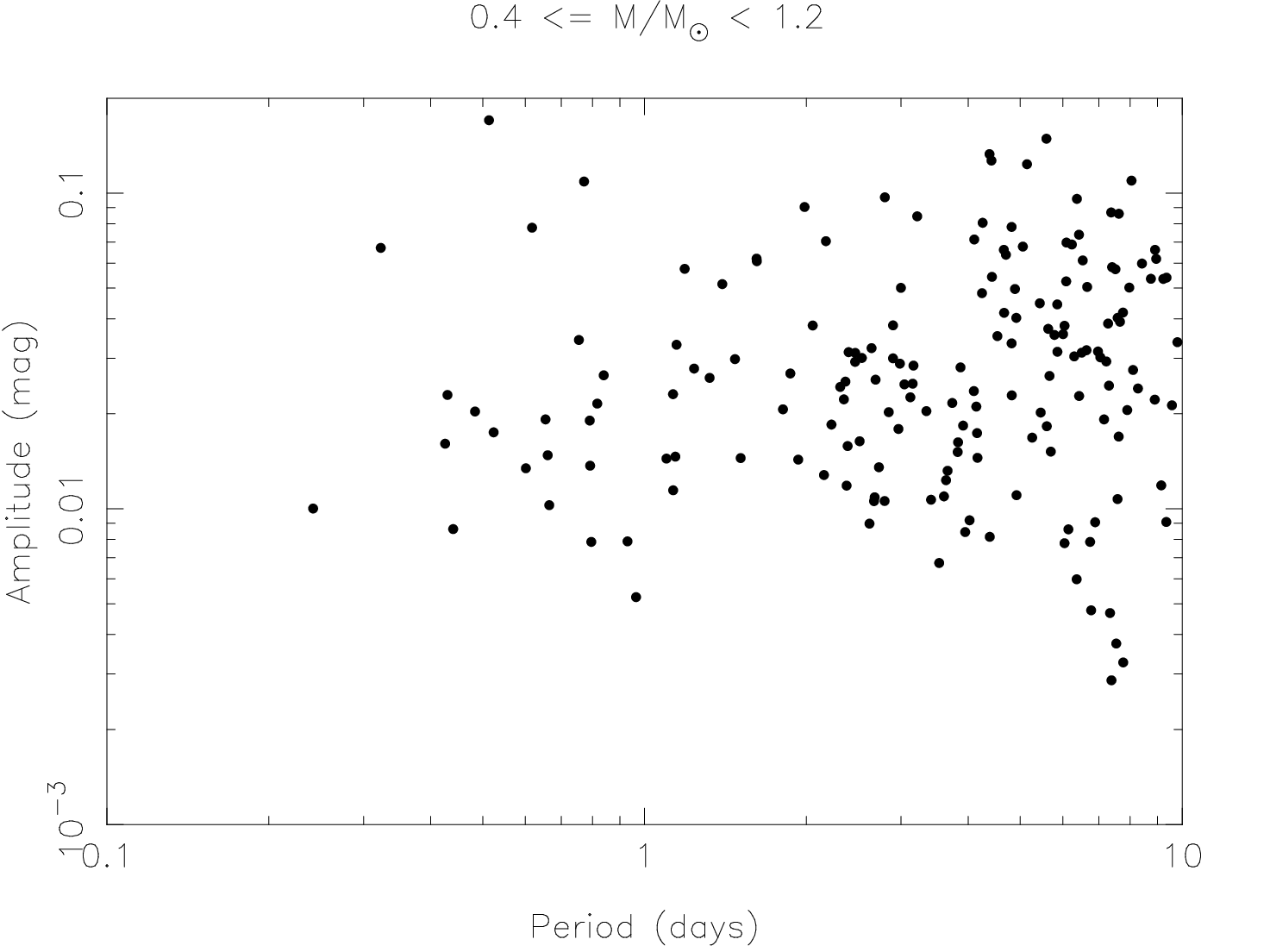}
\includegraphics[angle=270,width=3in]{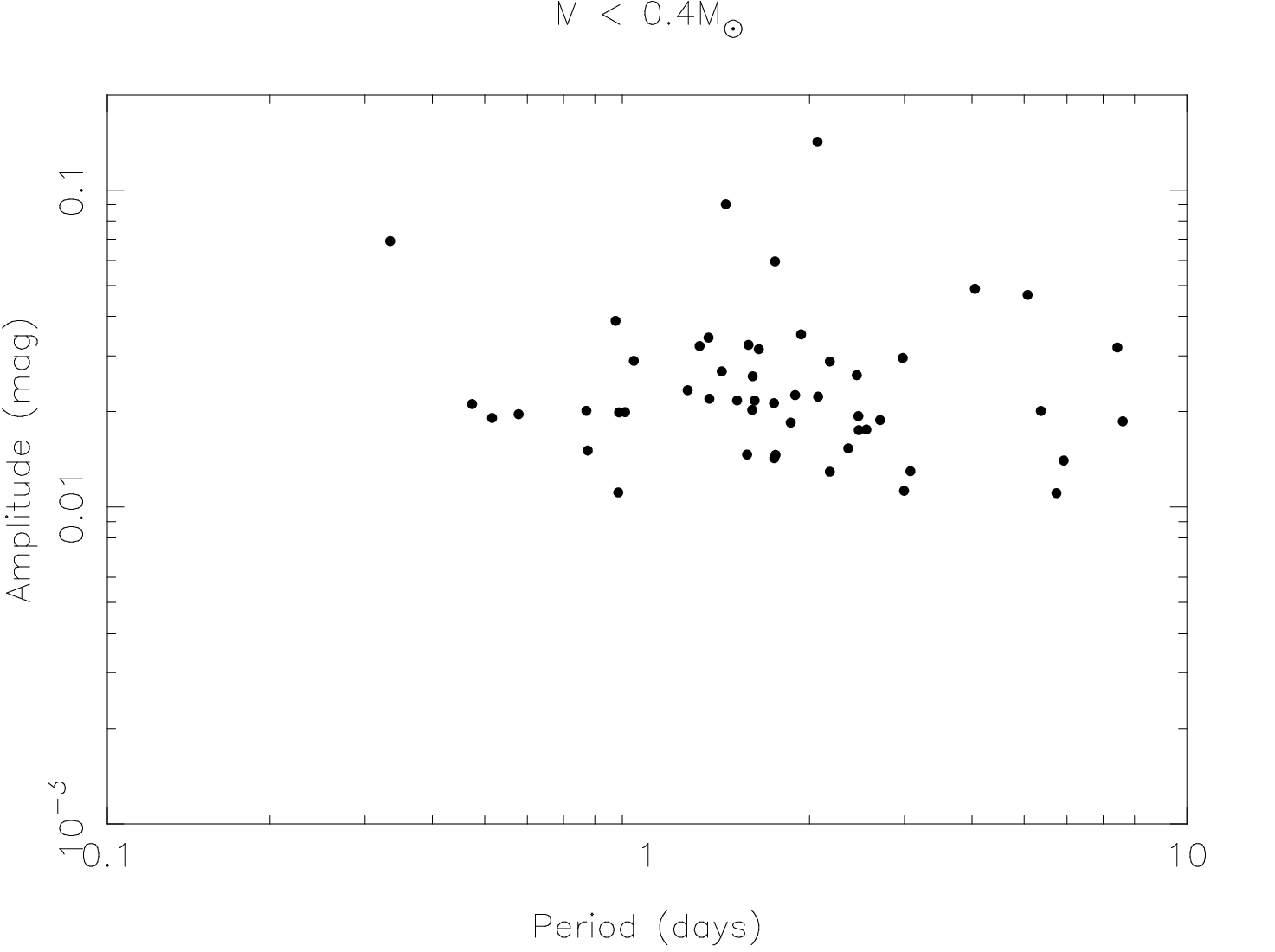}

\caption{Plot of amplitude as a function of period for NGC 2362 in two
  mass bins: $0.4 \le M/\msun < 1.2$ (top) and $M < 0.4\ \msun$
  (bottom).}

\label{pad}
\end{figure}

Figure \ref{pad} may indicate a slight bias toward shorter periods at
low masses, if the distribution of amplitudes in the  $0.4 \le M/\msun
< 1.2$ (top) and $M < 0.4\ \msun$ bins was the same, with no
counterpart to the small 
group of objects at periods of $\sim 5-10\ {\rm days}$ and amplitudes
$< 1$ per cent visible in the lower panel.  However, the result is not
highly significant due to small number statistics in the lower mass
bin, and it is possible that these objects may be false positive
detections, since at these amplitudes and periods, the effect of light
curve systematics is sometimes difficult to distinguish from real
variability.

\subsubsection{Period histograms}
\label{perioddist_section}

\begin{figure*}
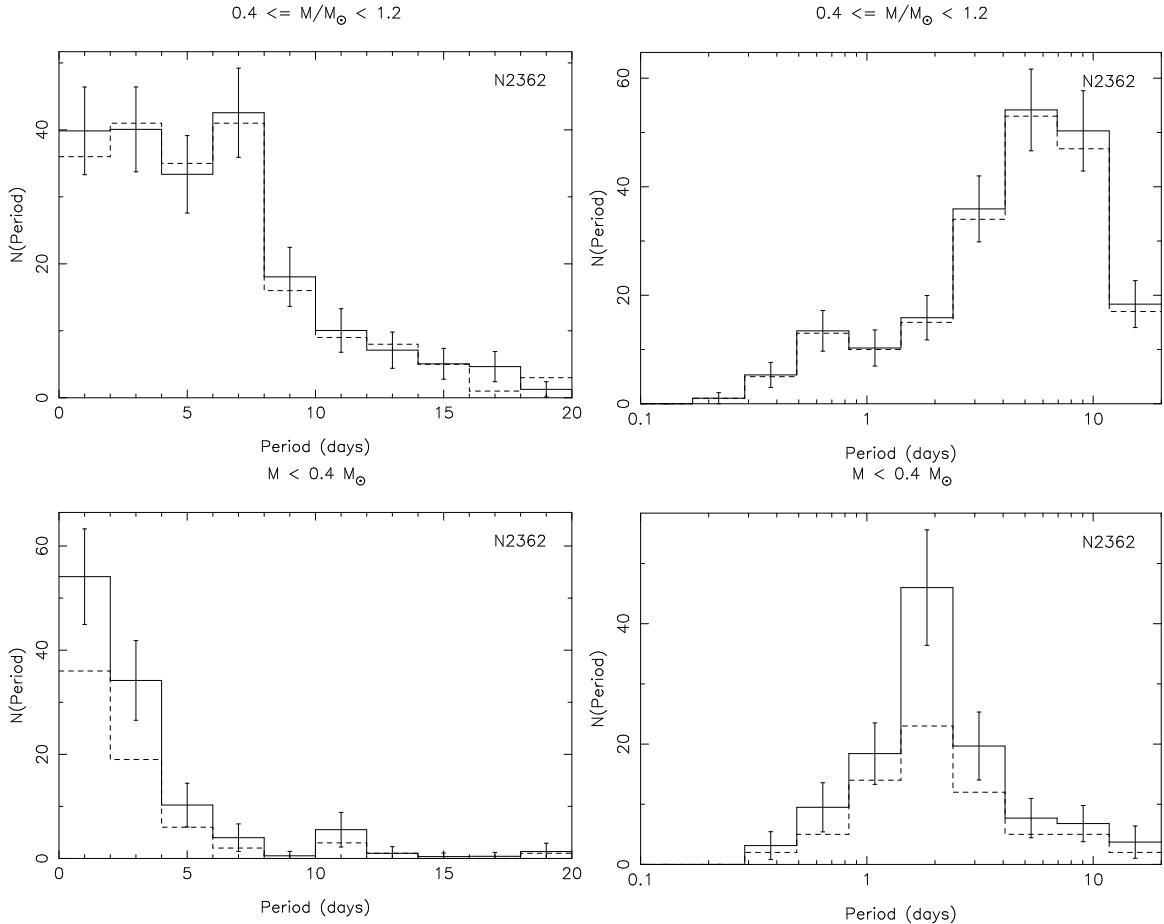

\centering
\includegraphics[angle=270,width=3in]{n2362_perioddist_1.ps}
\includegraphics[angle=270,width=3in]{n2362_perioddist_3.ps}
\includegraphics[angle=270,width=3in]{n2362_perioddist_2.ps}
\includegraphics[angle=270,width=3in]{n2362_perioddist_4.ps}

\caption{Period distributions for objects classified as possible
  photometric members, in two mass bins: $0.4 \le M/\msun < 1.2$
  (upper row, corresponding roughly to K and early-M spectral types)
  and $M < 0.4\ \msun$ (lower row, late-M).  The left-hand panels
  show the distributions plotted in linear period, and the right-hand
  panels show the same distributions plotted in $\log_{10}$ period.
  The dashed lines show the measured period distributions, and the
  solid lines show the results of attempting to correct for
  incompleteness and reliability, as described in the text.}

\label{perioddist}
\end{figure*}

In order to quantify the morphology of Figure \ref{pcd}, we have used
histograms of the rotation period distributions in two broad mass bins,
$0.4 \le M/\msun < 1.2$ and $M < 0.4\ \msun$, shown in Figure
\ref{perioddist}.  We have attempted to correct the distributions for
the effects of incompleteness and (un)reliability using the
simulations described in \S \ref{sim_section}, following the method
used in \citet{i2006}.  The results of doing this are shown in the
solid histograms in Figure \ref{perioddist}, and the raw period
distributions in the dashed histograms.

The period distributions in the two mass bins of Figure
\ref{perioddist} show clear differences, with the low-mass stars ($M <
0.4\ \msun$) showing a strongly peaked rotational period distribution,
with a maximum at $\sim 0.6-0.7\ {\rm days}$, whereas the higher-mass
stars ($0.4 \le M/\msun < 1.2$) show a broader distribution.  We
applied a two-sided Kolmogorov-Smirnov test to the corrected
distributions to confirm the statistical significance of this result,
finding a probability of $4 \times 10^{-10}$ that the distributions
were drawn from the same parent population.

The implication of this result is that the observed morphology in
Figure \ref{pcd}, and in particular the increase of the longest
observed rotation period as a function of increasing mass, a trend
followed also by the bulk of the rotators, is real and statistically
significant.

\subsection{Comparison with other data-sets}

\subsubsection{Period versus mass diagrams}
\label{pmd_section}

\begin{figure}
\centering
\includegraphics[angle=270,width=3in]{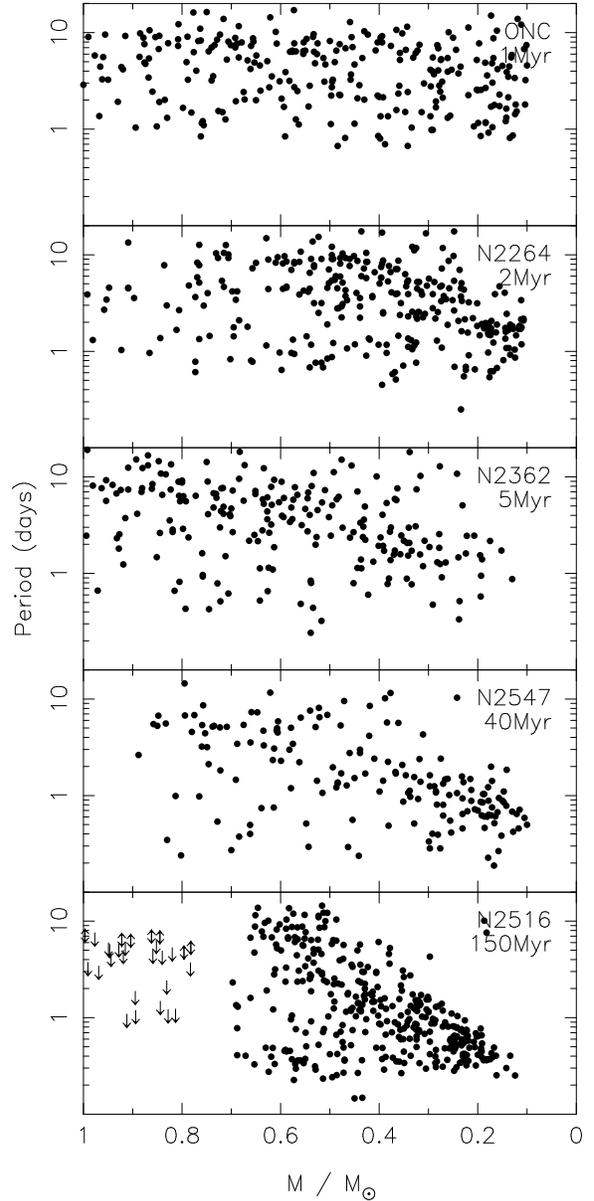}

\caption{Rotation period as a function of stellar mass for (top to
  bottom): ONC, NGC 2264, NGC 2362, NGC 2547 and NGC 2516.
  Lower and upper limits (from $v \sin i$ data) are marked 
  with arrows.  The masses were taken from the NextGen mass-magnitude
  relations \citep{bcah98} at the appropriate ages.  The ONC data are
  from \citet{h2002}.  For NGC 2264 we used the data of \citet{lamm05}
  and \citet{m2004}.  The NGC 2547 and NGC 2516 rotation period data
  are from the Monitor project (\citealt{i2007b}; \citealt{i2007d}).}

\label{pmd}
\end{figure}

Figure \ref{pmd} shows a diagram of rotation period as a function of
stellar mass for the ONC, NGC 2264, NGC 2362, NGC 2547 ($\sim
40\ {\rm Myr}$; \citealt{nj2006}) and NGC 2516 ($\sim 150\ {\rm Myr}$;
\citealt*{jth2001}).  Data sources for each cluster are indicated in
the figure caption.

The diagram clearly shows a gradual evolutionary sequence, from a
relatively flat mass-dependence of the rotation periods in the ONC
($\sim 1\ {\rm Myr}$), to a sloping relation in NGC 2264 and NGC 2362,
and an increasingly more pronounced slope at the older ages of NGC
2547 and NGC 2516.

The distribution for NGC 2362 appears to show a slightly more
pronounced sloping relation than NGC 2264, but the distributions are
very similar, indicating that these clusters are close in ``rotational
age''.  Nevertheless, the rotation period results seem to indicate that
the canonical age sequence, from youngest to oldest, of the ONC, NGC
2264 and then NGC 2362, is correct.

\subsubsection{Comparison with rotational evolution models}
\label{model_section}

By comparing the observed rotation periods as a function of age with
models of rotational evolution on the pre-main sequence, we can
constrain the amount of angular momentum loss required during this
part of the evolution.

In order to do this, we have used a simple model of solid body
rotation from \citet{i2007b}, including angular momentum losses due to
main sequence magnetically driven stellar winds (the effect of which
is close to negligible during this age range), but specifically, not
including any additional sources of angular momentum loss (``disc
locking'' or similar) in order to see if the latter are required to
explain the observations.  Our models from the earlier NGC 2516 and
NGC 2547 work incorporated a simple prescription for core-envelope
decoupling, but at these early ages the stars are fully-convective, so
it is not necessary to incorporate this effect here.

\begin{figure}
\centering
\includegraphics[angle=270,width=3in]{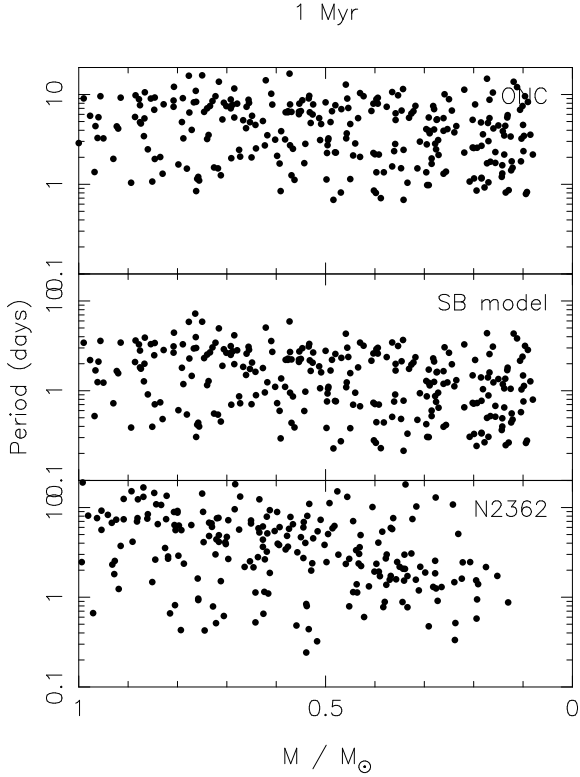}

\caption{Rotation period as a function of mass, using the model
  presented in \S \ref{model_section} to evolve the ONC
  distribution (top panel) forward in time from $1\ {\rm Myr}$ to $5\
  {\rm Myr}$ (middle panel), and the observed NGC 2362 distribution
  for comparison (bottom panel).  Note that these models assume no
  additional sources of angular momentum loss on the early-PMS.}

\label{pmd_evol_onc}
\end{figure}

Figure \ref{pmd_evol_onc} shows the results of attempting to evolve
the measured rotation periods in the ONC forward in time to the age
of NGC 2362 using these models.

The model clearly cannot successfully reproduce the NGC 2362
distribution.  Although there is some hint of the sloping relation at
low masses visible in the evolved ONC distribution, the slowest
rotators at $M \ga 0.6\ \msun$ are clearly rotating too rapidly.
However, the lower envelope of fast rotators is reasonably
well-reproduced throughout the mass range.

We suspect that the mass determinations (which are based on $I$-band
luminosity) are somewhat unreliable in the ONC, due to the presence of
strong spatially-variable extinction in this cluster.  This introduces
horizontal scatter in the period versus mass plane.  It has been
necessary to assume a typical (low) value of this parameter for many
of the low-mass detections due to a lack of suitable estimates based 
on spectroscopy from \citet{h97}.  Hence, much of the scatter at low
masses in the figure may be due to underestimation of the extinction,
and hence of the masses, for highly reddened objects.

However, at the high-mass end, the diagram is more reliable, since the
majority of the scatter due to extinction will only move objects to
the right.  Therefore, the result that the slowest rotators at these
masses are rotating too rapidly is still significant.  This indicates
that an additional source of angular momentum loss is required, in
order to hold the rotation periods of these stars approximately
constant (since the slowest rotators in the ONC and NGC 2362 appear to
have very similar periods of $\sim 10\ {\rm days}$).  This is
precisely the result found by many other authors, and provides
evidence in support of disc regulation, or a similar mechanism,
operating over the early-PMS stages of the evolution.

\begin{figure}
\centering
\includegraphics[angle=270,width=3in]{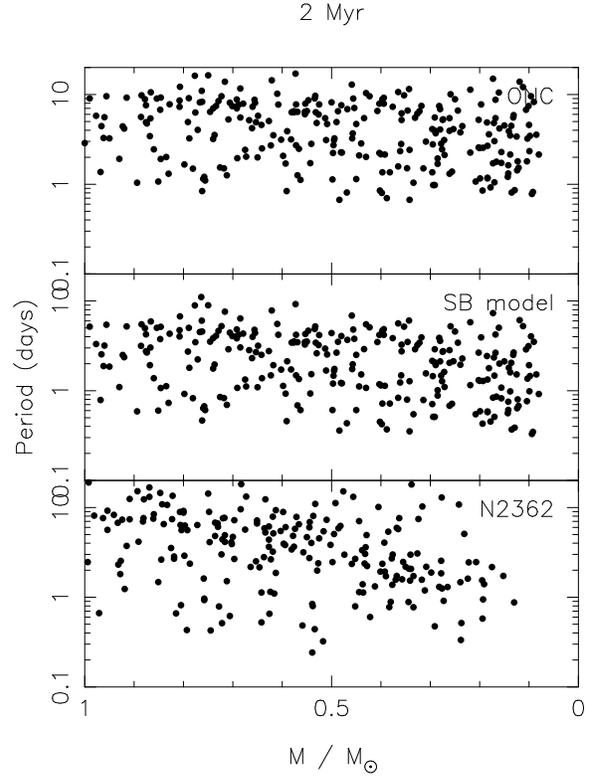}

\caption{As Figure \ref{pmd_evol_onc}, except assuming a $2\ {\rm
    Myr}$ age for the ONC.}

\label{pmd_evol_onc_2}
\end{figure}

We note that the age of the ONC is still a subject of controversy in
the literature at the time of writing.  Some recent studies
(e.g. \citealt{sand07}; \citealt{j2007}) have indicated that the
distance to the ONC may be $\sim 390\ {\rm pc}$, rather than the
conventional value of $470\ {\rm pc}$ used by \citet{h97}.  Depending
on the PMS evolutionary tracks used, this can raise the derived age to
$\sim 2\ {\rm Myr}$, which may help to explain these apparent
discrepancies.  We have explored this further in Figure
\ref{pmd_evol_onc_2}, which shows the result of assuming a $2\ {\rm
  Myr}$ age for the ONC, and evolving the rotation period distribution
forward in time.  Since the models then predict less spin-up, it is
not surprising that the slow rotators at solar mass are better-fit,
but still with evidence for (more modest) angular momentum 
loss being required to fully-reproduce the NGC 2362 slow rotator
population.  However, at lower masses, the resulting slower
rotation for the bulk of these objects places the upper limit to the
rotation periods even higher than observed in NGC 2362.  Raising the
age of the ONC does not therefore readily provide an explanation for
the problems at the low-mass end that have already been discussed.

\begin{figure}
\centering
\includegraphics[angle=270,width=3in]{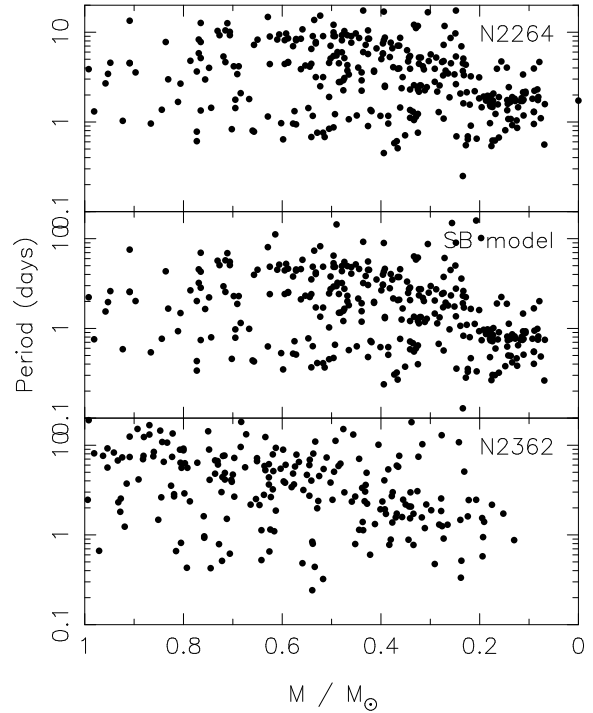}

\caption{As Figure \ref{pmd_evol_onc}, except using NGC 2264 as the
  initial condition, assuming an age of $2\ {\rm Myr}$.}

\label{pmd_evol_n2264}
\end{figure}

We have also repeated the analysis using NGC 2264 as the initial
condition.  Figure \ref{pmd_evol_n2264} shows the results, where we
have assumed a $2\ {\rm Myr}$ age for this cluster.  NGC 2264 has much
less differentially-variable reddening than the ONC, so the low-mass
end of the diagram should be more reliably reproduced in this
cluster.

The diagram indicates that this is indeed the case, and that the
sloping relation for the slow rotators at low masses in NGC 2362 is
well-reproduced by evolving the NGC 2264 observations forward in
time.  Again however, the slowest rotators at higher masses ($M \ga
0.6\ \msun$) appear to rotate too rapidly compared to the NGC 2362
observations.

Given that these models do not incorporate additional sources of
angular momentum loss applying to the early-PMS, such as disc
regulation, the ability of the models to reproduce the form of the
sloping relation for the lowest-mass stars ($M \la 0.5\ \msun$)
indicates that these stars do not experience significant angular
momentum loss at this stage of their evolution, in contrast to the
higher-mass stars.

We also demonstrated in our earlier NGC 2516 and NGC 2547 publications
(\citealt{i2007b}; \citealt{i2007d}), that a similar statement
applies to the lowest-mass stars at later stages in their evolution:
even from the $\sim 5\ {\rm Myr}$ age of NGC 2362 to the $\sim 150\
{\rm Myr}$ age of NGC 2516, additional angular momentum loss at $M \la
0.5\ \msun$ above that expected due to (saturated) stellar winds does
not appear to be required.  Given that the higher-mass stars do appear
to need additional angular momentum losses, this implies that such
losses are mass-dependent, in the sense that lower-mass stars
experience less PMS angular momentum loss, or that they lose angular
momentum over a shorter timescale ($\ll 5\ {\rm Myr}$) than the
higher-mass stars.

\subsection{Comparison with mid-IR \spitzer observations}
\label{spitzer_sect}

In order to search for the correlation between rotation period and the
presence of discs expected in the disc regulation paradigm for
early-PMS angular momentum evolution, we combined the present optical
data-set with the \spitzer IRAC photometry of \citet{dh07}.  Note that
\citet{dh07} only give mid-IR measurements for objects detected in
their combined photometric and spectroscopic survey of the central
$\sim 11' \times 11'$ of the cluster \citep{dahm05}, whereas the
\spitzer field-of-view is somewhat larger, so this decreases the
available sample size.  A re-analysis of the \spitzer observations
to resolve this problem will be attempted in a future publication.

There are $64$ stars in common with the catalogue of \citet{dh07}
which have rotation period measurements.  However, in order to obtain
a meaningful result, we must restrict the mass range over which the
samples are compared: not doing so would introduce a bias at the faint
end toward those objects with excesses, since they are more readily
detected in the mid-IR data.  Examining Figure 1 of \citet{dh07}
indicates that the IRAC detection limit is at $[4.5] \simeq 15$, and
using the models of \citet{bcah98} for the $M$-band (which is a
reasonable approximation to the \spitzer $4.5\ \micron$ band), this
corresponds to a conservative limit of $M > 0.4\ \msun$, which has
been adopted henceforth.

\begin{figure}
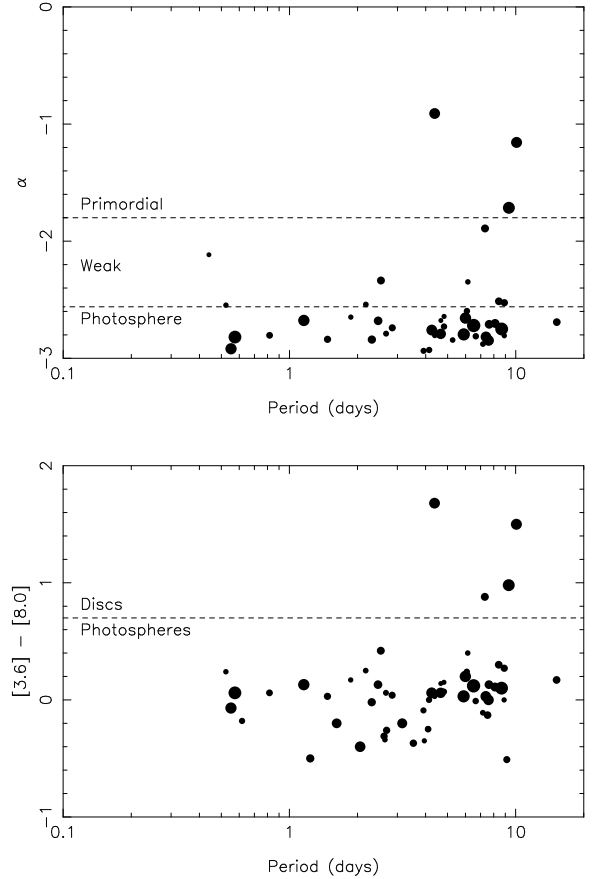

\centering
\includegraphics[angle=270,width=3in]{n2362_alpha_prot.ps}
\includegraphics[angle=270,width=3in]{n2362_irex_prot.ps}

\caption{Two disc indicators plotted as a function of rotation
  period.  Symbol sizes indicate the object masses, from $0.4\ \msun$
  (smallest symbols) to $1.2\ \msun$ (largest symbols).  {\bf Top:}
  SED slope $\alpha$ as a function of rotation period.  {\bf Bottom:}
  $[3.6] - [8.0]$ colour as a function of rotation period.  The dashed
  lines indicate the approximate boundaries in these quantities
  between regions of the diagram occupied by objects with and without
  discs, as discussed in the text.}

\label{disc_period}
\end{figure}

We have considered two commonly-used indicators of the presence of a
circumstellar disc: the slope of the IRAC spectral energy distribution
(SED), $\alpha = d\log(\lambda F_\lambda) / d\log \lambda$, as used by
\citet{dh07}, and the $[3.6] - [8.0]$ colour, as used by \citet{cb06}
and \citet{cb07}.  Figure \ref{disc_period} shows a plot of these
quantities as a function of rotation period.

Thresholds were defined in these quantities, following \citet{dh07}
for $\alpha$, with stellar photospheres having $\alpha < -2.56$,
``weak discs'' $-2.56 < \alpha < -1.80$, and ``primordial'' or
optically thick discs $\alpha \ge -1.80$.  In $[3.6] - [8.0]$ colour,
we follow \citet{cb07}, and use a threshold of $[3.6] - [8.0] > 0.7$
to identify objects with mid-IR excesses indicative of the presence of
discs.  Figure \ref{disc_period} clearly indicates that the
``primordial'' discs as defined by SED slope, and the excess objects
in $[3.6] - [8.0]$, are essentially the same.

\begin{figure}
\centering
\includegraphics[angle=270,width=3in]{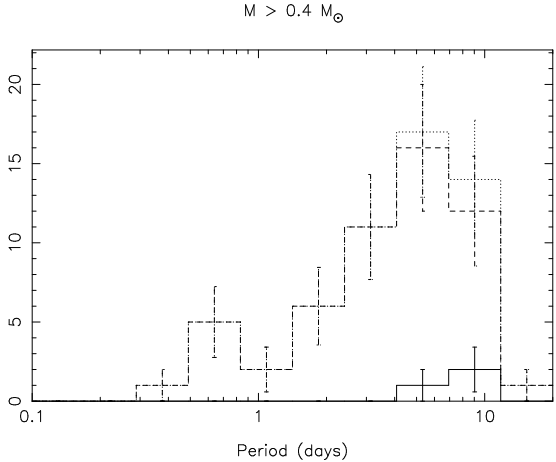}

\caption{Histograms of rotation period for $M > 0.4\ \msun$, showing:
  objects with IR excesses (``primordial discs'', or $\alpha \ge
  -1.80$: solid line), objects without large IR excesses (``weak
  discs'' and ``photospheres'', or $\alpha < -1.80$: dashed line), and
  the entire sample for comparison (dotted line).  Poisson error bars
  are shown.}

\label{disc_phist}
\end{figure}

Both panels of Figure \ref{disc_period} indicate that the $3-4$
objects showing strong disc signatures all have long periods, whereas
the remainder of the objects show a range of periods, as would be
expected in the disc regulation paradigm.  This is also seen in
distributions of rotation periods in these bins, which are shown in
Figure \ref{disc_phist}.

Applying a two-sided K-S test to the distributions of objects with and
without IR excesses gave probabilities of $0.12$ (using $\alpha \ge
-1.80$ to select the objects with significant excesses, corresponding
to the ``primordial disc'' objects in Figure \ref{disc_period}), and
$0.15$ (using $[3.6] - [8.0] > 0.7$) that the distributions were drawn
from the same parent population.  Since the K-S test is not strictly
appropriate for such a small sample size, we also considered a simple
probabilistic argument: of the $55$ objects without excesses, $30$
have periods $> 4\ {\rm days}$.  The probability that all three
objects with excesses were drawn from this same parent population is
therefore $0.16$.  These results indicate that the differences are not
statistically significant.  This is most likely due to the small
sample size.

In spite of this, the distributions do indeed hint at the expected
result for the disc regulation paradigm, that the objects with discs
should be slow rotators, as seen e.g. by \citet{rebull06} and
\citet{cb07}.  Obtaining a larger sample of objects with rotation
period and mid-IR measurements should improve the significance of
these results, and allow more definitive conclusions to be made.

\section{Conclusions}
\label{conclusions_section}

We have reported on results of an $i$-band photometric survey of NGC
2362, covering $\sim 0.4\ {\rm sq. deg}$ of the cluster.  Selection
of candidate members in a $V$ versus $V-I$ colour-magnitude diagram
using an empirical fit to the cluster sequence found $1800$
candidate members, over a $V$ magnitude range of $16 < V < 26$
(covering masses from $1.2\ \msun$ down to below the brown dwarf
limit).  The likely field contamination level was estimated using a
simulated catalogue of field objects from the Besan\c{c}on Galactic
models \citep{r2003}, finding an overall contamination level of $\sim
65$ per cent, implying that there are $\sim 630$ real cluster members
over this mass range in our field-of-view.

We derived light curves for $\sim 85\,000$ objects in the NGC 2362
field, achieving a precision of $< 1$ per cent per data point over $15
\la i \la 19$.  The light curves of our candidate cluster members were
searched for periodic variability corresponding to stellar rotation,
giving $271$ detections over the mass range $0.1 < M/\msun < 1.2$.

The rotation period distribution as a function of mass was found to
show a clear mass-dependent morphology, as seen in our earlier NGC
2547, NGC 2516 and M34 studies, and qualitatively similar to that in
NGC 2264, which is thought to have a similar age.

Using rotational evolution models from our earlier NGC 2516 work
\citep{i2007b}, we were able to show that distribution of rotation
rates for the rapid rotators in NGC 2362 could be reproduced from the
distributions in the ONC and NGC 2264 by evolving them forward in time
to $\sim 5\ {\rm Myr}$.  However, for the slow rotators at masses $\ga
0.6\ \msun$, the predicted rotation rates from the models were too
rapid, with the observations indicating that the rotation periods 
of these stars are in fact approximately constant over the $\sim 1-5\
{\rm Myr}$ age range.  This confirms the need for additional angular
momentum losses on this part of the PMS, e.g. as would be expected in
the disc regulation paradigm.

Finally, we have used the \spitzer observations of \citet{dh07} to
test the disc regulation hypothesis, by searching for a correlation
between rotation rate and mid-IR excesses indicative of circumstellar
discs.  No statistically significant correlation was found with the
present small sample sizes, but the $3-4$ objects with significant
disc signatures in the sample are indeed all slow rotators.

\section*{Acknowledgments}

Based on observations obtained at Cerro Tololo Inter-American
Observatory, a division of the National Optical Astronomy
Observatories, which is operated by the Association of Universities
for Research in Astronomy, Inc. under cooperative agreement with the
National Science Foundation.  This publication makes use of data
products from the Two Micron All Sky Survey, which is a joint project
of the University of Massachusetts and the Infrared Processing and
Analysis Center/California Institute of Technology, funded by the
National Aeronautics and Space Administration and the National Science
Foundation.  This research has also made use of the SIMBAD database,
operated at CDS, Strasbourg, France, and the WEBDA database, operated
at the Institute for Astronomy of the University of Vienna.  The Open
Cluster Database, as provided by C.F. Prosser and J.R. Stauffer, may
currently be accessed at {\tt
  http://www.noao.edu/noao/staff/cprosser/}, or by anonymous ftp to
{\tt 140.252.1.11}, {\tt cd /pub/prosser/clusters/}.

JI gratefully acknowledges the support of a PPARC studentship, and SA
the support of a PPARC postdoctoral fellowship.  We would like to
express our gratitude to Isabelle Baraffe for providing the stellar
evolution model tracks used in \S \ref{model_section}, Scott Dahm for
providing the catalogue of \spitzer observations used in \S
\ref{spitzer_sect}, and Aleks Scholz for helpful advice on the
handling of \spitzer data.  We also thank the referee for his
comments, which have helped to improve the manuscript.

\end{document}